\documentclass[twocolumn,showpacs,preprintnumbers,amsmath,amssymb, superscriptaddress]{revtex4-1}
\usepackage[toc,page]{appendix}
\usepackage{graphicx}
\usepackage{bm}

\begin{document}

\title{The critical phenomena of a single defect}

\author{Xintian Wu}
\email{Corresponding author. Email: wuxt@bnu.edu.cn}
\author{Yangyang Zhang}

\affiliation{Department of Physics, Beijing Normal University, Beijing, 100875, China}

\date{\today}

\begin{abstract}
 We consider the critical system with a point defect and study the variation of thermodynamic quantities, which are the differences between those with and without the defect.  Within renormalization group theory, we show generally that the critical exponent of the internal energy variation is the specific heat exponent of a pure system, and the critical exponent of the heat capacity variation is that for the temperature derivative of specific heat of a pure system. This conclusion is valid for the isotropic systems with a short-range interaction. As an example we solve the two dimensional Ising model with  a point defect numerically. The variations of the free energy, internal energy and specific heat are calculated with bond propagation algorithm. At the critical point, the internal energy variation diverges with the lattice size logarithmically and  the heat capacity variation  diverges with size linearly. Near the critical point, the internal energy variation behaves as $\ln |t|$ and the heat capacity variation behaves as $|t|^{-1}$, where $t$ is the reduced temperature. 
\end{abstract}

\pacs{75.10.Nr,02.70.-c, 05.50.+q, 75.10.Hk}

\maketitle

\section{Introduction}

As we know there are many dramatic effects in the critical phenomena because of the divergence of the correlation length at the critical point \cite{stanley}. For example, the specific heat has a logarithmic divergence for the two dimensional Ising model \cite{onsager}. Are there some dramatic effects near the critical point when we add a point defect into the system? The answer is yes. The effects are dramatic and surprising.

Consider the two dimensional Ising model with a point defect. We focus on the variations, the difference between with and without the defect, of thermodynamic quantities. Naively, one would expect the internal energy variation $\delta U \propto u_0$, where $u_0$ is the energy density of pure system and is a constant at the critical point. {\bf However in our numerical calculation on a $N \times N$ lattice,  at the critical point the internal energy variation is proportional to $\log N$. Moreover the heat capacity variation is proportional to $N$ rather than $\log N$}. Hence the defect makes an unexpectedly large contribution.   

Using renormalization group (RG) theory, we studied the influence of the defect on the thermodynamic quantities. Due to the long range correlation near the critical point, the defect changes the energy density across the whole system. We find that the internal energy variation is an integration of energy correlation, which gives rise to the specific heat. Then the internal energy  variation is proportional to the specific heat of the pure system. Consequently the heat capacity variation is proportional to the temperature derivative of specific heat of a pure system.  With operator product expansion (OPE) \cite{kadanoff,cardy} and conformal field theory (CFT) \cite{cardy1,francesco}, we show these conclusions for the two dimensional Ising model explicitly. Then we use the scaling theory and RG to extend to common critical systems.

To test the above conclusions, we study the two dimensional Ising model on a finite size lattice with a site defect using bond propagation algorithm (BPA) \cite{loh1,loh2,wu1,wu2,wu3}. The variations induced by the defect of the free energy, internal energy and heat capacity are calculated. The BPA results verify the RG and scaling argument about the internal energy and heat capacity variation. 

In fact, the effect of defects on the critical phenomena has been studied 40 years ago \cite{osawa,harris,shafer}. However those studies focused on the average effect of many defects rather than a single defect. In addition, this study is not purely academic, but has potential practical use.  The rapid development of Nano techniques  makes it possible to study the critical phenomena on small size systems \cite{kawata,puntes,yin}. In the production of the samples, the defect is unavoidable usually. Therefore it is of interest to study the effect of a defect in a finite size  system.

We arrange this paper as follows. In the section II, we present RG and scaling argument. In section III, we report the BPA results on the two dimensional Ising model with a defect. Section IV is a concluding remark and acknowledgement.

\section{The critical exponents of the internal energy and heat capacity variation}

We shall prove generally that for a spin system with short range interaction in any dimension:

1. The critical exponent of the internal energy variation due to a site  defect is the same as that of the specific heat of a pure system.
 
2. The critical exponent of the heat capacity variation  is the same as that of the temperature derivative of specific heat of a pure system.

At first we show these conclusions for a two dimensional Ising model explicitly. Then we give a general scaling argument for common critical systems.

\begin{figure}
\includegraphics[width=0.5\textwidth]{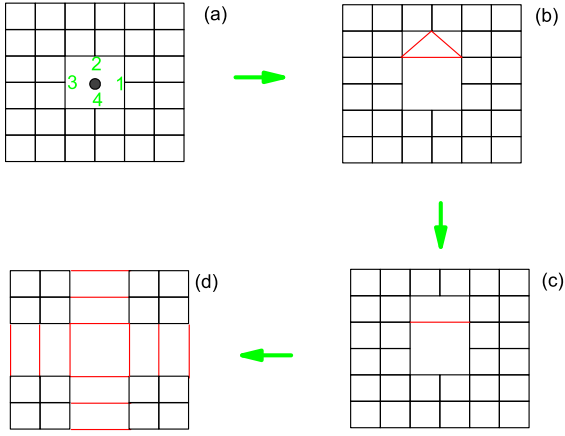}
\caption{ (a) $7 \times 7$ lattice with a defect in the centre. The spin $0$ is labelled by the black spot at the center. Its four neighbors are labeled by $1,2,3,4$. The bonds between spin $0$ and $1,2,3,4$ are zero. All other bonds are equal to $1$. (b) After applying $Y-\Delta$ transformation to the four spins in (a), where the center spin is spin $2$, we get this lattice. (c)Using BPA, we get this lattice. (d) Applying the same procedure in three other directions, the original $7\times 7$ lattice is transformed into a $6\times 6$ lattice.  }
\label{boundary}
\end{figure}

Consider the two dimensional Ising model on a square lattice with a site defect. The Hamiltonian is given by
\begin{equation}
H=-\sum_{<i,i'>} J_{ii'}\sigma_i \sigma_{i'}
\end{equation}
with $J_{01}=J_{02}=J_{03}=J_{04}=0$ and all other bonds $J_{ii'}=1$. The summation is over the nearest neighbours. A $7\times 7$ such lattice is shown in the Fig. 1. We will study this model with BPA in the next section, so we also show the schematic of the BPA.   Its partition function is given by
\begin{equation}
Z=Tre^{-\beta H}\equiv \sum_{\{\sigma_i\}}e^{-\beta H}
\end{equation}

The Hamiltonian for the  pure system is given by
\begin{equation}
H_0=-\sum_{<i,i'>} \sigma_i \sigma_{i'}.
\end{equation}
and its partition function is given by
\begin{equation}
Z_0=Tre^{-\beta H_0}\equiv \sum_{\{\sigma_i\}}e^{-\beta H_0}
\end{equation}

Define the variation of the Hamiltonian
\begin{equation}
\delta H=H-H_0=\sum_{i=1}^4\sigma_0\sigma_i.
\end{equation}
To study the effect of the defect, we will calculate the variation of free energy defined by
\begin{equation}
\delta F=F_0-F
\end{equation}
where $ F=-\ln Z$ and $F_0=-\ln Z_0$; the internal energy variation
\begin{equation}
\delta U=U_0-U
\end{equation}
where $ U= \partial F /\partial \beta$ and $U_0=\partial F_0 /\partial \beta$;  and the heat capacity variation
\begin{equation}
\delta C=C_0-C
\end{equation}
where $ C= \partial^2 F /\partial \beta^2$ and $C_0=\partial^2 F_0 /\partial \beta^2$.

\subsection{ The RG and CFT argument for two dimensional Ising model with a defect}

For the free energy variation, we have
\begin{equation}
\delta F=\ln \frac{Tr(e^{-\beta (H_0+\delta H)})}{Tr(e^{-\beta H_0})}=\ln \left \langle e^{-\beta \delta H}\right\rangle _0
\end{equation}
Here we introduce the notation $\left\langle \cdots\right\rangle _0 =Tr[\cdots e^{-\beta H_0}]/Tre^{-\beta H_0}$.
Using the equation
\begin{equation}
e^{-\beta  \sigma_i \sigma_j}\equiv X-Y\sigma_i \sigma_j,
\label{eq:twoterm}
\end{equation}
where $X=\cosh \beta, Y=\sinh \beta$, we can expand $e^{-\beta \delta H}$ into
\begin{eqnarray}
& & e^{-\beta \delta H} = \prod_{i=1}^{4} (X-Y\sigma_0\sigma_i) \nonumber \\
& = & A_1-A_2\sum_{i=1}^{4}\sigma_0\sigma_i+A_3\sum_{i<j}^{4}\sigma_i\sigma_j \nonumber \\
& & -A_4\sum_{i<j<k}^{4}\sigma_0\sigma_i\sigma_j\sigma_k+A_5\sigma_1\sigma_2\sigma_3\sigma_4  \nonumber \\
\label{eq:expansion}
\end{eqnarray}
where the coefficients are given by $A_1=X^4,A_2=X^3Y,A_3=X^2 Y^2,A_4=XY^3,A_5=Y^4$. The free energy variation is given by the average of these local variables.  It is singular since $F_0$ is singular. We will show the singularity of the internal energy (derivative of $\delta F$) variation explicitly.

For the internal energy variation we have
\begin{equation}
U_0= \frac{Tr(H_0e^{-\beta H_0} )}{Tr(e^{-\beta H_0})}\equiv \left\langle H_0\right\rangle _0
\end{equation}
and
\begin{equation}
U = \frac{\left\langle e^{-\beta \delta H}(H_0+\delta H)\right\rangle _0}{\left\langle e^{-\beta \delta H}\right\rangle _0}.
\end{equation}
 Then we get
\begin{equation}
\delta U=\frac{\left\langle e^{-\beta \delta H}(H_0+\delta H)\right\rangle _0}{\left\langle e^{-\beta \delta H}\right\rangle _0}-\left\langle H_0\right\rangle _0.
\label{eq:delta-u}
\end{equation}

Substituting equation (\ref{eq:expansion}) into $\delta U$, we get an expansion. Because $\left\langle e^{-\beta \delta H}\delta H \right\rangle _0/\left \langle e^{-\beta \delta H} \right \rangle _0$ is a higher order contribution as compared to $\left\langle e^{-\beta \delta H}H_0\right\rangle _0/\left\langle e^{-\beta \delta H}\right\rangle _0$  , we ignore it. The contribution to $\delta U$ from the second term in Eq. (\ref{eq:expansion}) is given by
\begin{equation}
\sum_{i=1}^{4} \sum_{<k,k'>}(\left\langle \sigma_0\sigma_i \sigma_k\sigma_{k'}\right\rangle _0-\left\langle \sigma_0\sigma_i\right\rangle _0\left\langle  \sigma_k\sigma_{k'}\right\rangle _0 ).
\label{eq:u-1}
\end{equation}
Here and below we neglect the denominator $\left\langle e^{-\beta \delta H}\right\rangle _0$, which is a finite quantity and can be replaced by its value at the critical point.  The quantity in the above equation is proportional to the specific heat of the pure system. The second non zero term in the expansion of $\delta U$ related to the third term in Eq. (\ref{eq:expansion}) is given by
\begin{equation}
\sum_{i<j}^{4} \sum_{<k,k'>}(\left\langle \sigma_i\sigma_j \sigma_k\sigma_{k'}\right\rangle _0-\left\langle \sigma_i\sigma_j\right\rangle _0\left\langle  \sigma_k\sigma_{k'}\right\rangle _0 ).
\label{eq:u-2}
\end{equation}

With OPE in RG we will show both terms in the above two equations are of the same order and proportional to the specific heat of a pure system. Generally, a product of two nearby
basic operators is reducible as \cite{kadanoff,cardy,cardy1,francesco}
\begin{eqnarray}
O_{\alpha}({\bf r}_1)O_{\alpha}({\bf r}_2) & = &\sum_{\gamma}c_{\alpha\beta,\gamma}({\bf r}_{12}) \phi_{\gamma}({\bf R}_{12}) \nonumber \\
 {\bf r}& = & {\bf r}_1-{\bf r}_2, \hskip 0.2cm {\bf R}=({\bf r}_1+{\bf r}_2)/2
\label{eq:ope}
\end{eqnarray}
where $c_{\alpha\beta,\gamma}$ are OPE coefficients. The sum is over all the scaling operators. The exact meaning of the above equation is defined by \cite{cardy1}
\begin{equation}
\langle O_{\alpha}({\bf r}_1)O_{\alpha}({\bf r}_2) \cdots \rangle_0 = \langle \sum_{\gamma}c_{\alpha\beta,\gamma}({\bf r}_{12}) \phi_{\gamma}({\bf R}_{12}) \cdots\rangle_0 
\end{equation}
The point is that, in the limit when $|{\bf r}_1 - {\bf r}_2 |$ is much less than the separation between ${\bf r}_1,{\bf r}_2 $ and all the other arguments in $\cdots$, the coefficients $c_{\alpha\beta,\gamma}$ are independent of what is in the dots.

For two nearby spin operators in two dimensional Ising model, we know the exact expansion from the conformal field theory (CFT) \cite{francesco}
\begin{equation}
\sigma_0 \sigma_1\sim |{\bf r}_{01}|^{-\frac{1}{4}}+\frac{1}{2}|{\bf r}_{01}|^{\frac{3}{4}}\epsilon({\bf R}_{01})+\cdots
\label{eq:cft-1}
\end{equation}
where $\epsilon$ is the energy density operator, and $\cdots$ are irrelevant operators with larger scaling dimensions. The energy operator $\epsilon ({\bf r})$ is also called thermal operator. Note that $\epsilon$ is the operator for energy density minus its critical value, so $\left\langle \epsilon({\bf r})\right\rangle _0=0$. Then $\sigma_i \sigma_j$ with $1\le i<j\le 4$ in Eq. (\ref{eq:u-2}) can also expanded as linear combination of identity and energy density operators. Both Eq. (\ref{eq:u-1}) and (\ref{eq:u-2}) give rise to energy correlators.
For a energy correlator,
\begin{equation}
\left\langle \epsilon(0)\epsilon({\bf r})\right\rangle _0\sim \frac{e^{-r/\xi_{\pm}}}{r^2}
\end{equation}
where $\xi_{\pm}$ is the correlation length for $t>0$ and $t<0$ respectively, and $t$ is the reduced temperature..

From Eq. (\ref{eq:cft-1}), we get their contribution of terms in Eq. (\ref{eq:u-1}), (\ref{eq:u-2}) to the internal energy variation
\begin{eqnarray}
&  & \sum_{i=1}^{4} \sum_{<k,k'>} \frac{1}{4}|{\bf r}_{0i}|^{\frac{3}{4}}|{\bf r}_{kk'}|^{\frac{3}{4}}\left\langle \epsilon({\bf R}_{0i})\epsilon({\bf R}_{kk'})\right\rangle _0  \nonumber \\
& + & \sum_{i<j}^{4} \sum_{<k,k'>} \frac{1}{4}|{\bf r}_{ij}|^{\frac{3}{4}}|{\bf r}_{kk'}|^{\frac{3}{4}}\left\langle \epsilon({\bf R}_{ij})\epsilon({\bf R}_{kk'})\right\rangle _0 \nonumber \\
& \sim & \frac{1}{4}\sum_{<k,k'>}|{\bf r}_{kk'}|^{\frac{3}{4}} (\sum_{i=1}^{4}|{\bf r}_{0i}|^{\frac{3}{4}}|+\sum_{i<j}^{4}|{\bf r}_{ij}|^{\frac{3}{4}})
\frac{e^{-r_k/\xi_{\pm}}}{r_k^2}\label{eq:u-2-a} \nonumber \\
\end{eqnarray}
Considering that $i,j$ are the neighbors of the defect and $k,k'$ are neighbors, the approximation $|{\bf R}_{kk'}-{\bf R}_{0i}|\approx |{\bf R}_{kk'}-{\bf R}_{ij}|\approx r_k$ is used, where $r_k$ is the separation between the spin $k$ and the defect. One may note that in the summation of Eq. (\ref{eq:u-1}), (\ref{eq:u-2}) there are terms $\left\langle \sigma_i\sigma_j \sigma_k\sigma_{k'}\right\rangle _0$ with the positions of $\sigma_{k}, \sigma_{k'}$  are not far from the defect. For such terms, we can not get the simple energy correlators in the above equations. However the main contribution comes from those terms with $\sigma_{k}, \sigma_{k'}$ far from the defect. For these terms, we can apply OPE to expand $\sigma_i\sigma_j$ and
$\sigma_k \sigma_{k'}$ with Eq. (\ref{eq:cft-1}).

The left two terms in Eq. (\ref{eq:expansion}) are products of four operators. For the fourth term in Eq. (\ref{eq:expansion}), we take the term $\sigma_0\sigma_1\sigma_2\sigma_3$ as an example. It can be expanded as
\begin{equation}
\sigma_0\sigma_1\sigma_2\sigma_3=c_{\sigma\sigma\sigma\sigma, I}+c_{\sigma\sigma\sigma\sigma, \epsilon}\epsilon({\bf R}_{0123})+\cdots.
\end{equation}
In this expansion, another relevant scaling operator $\sigma$ does not exist because it has different symmetry with the product of four spin operators. The coefficients $c_{\sigma\sigma\sigma\sigma, I}$,  $c_{\sigma\sigma\sigma\sigma, \epsilon}$  are functions of ${\bf r}_{0},{\bf r}_{1},{\bf r}_{2},{\bf r}_{3}$. The position ${\bf R}_{0123}=( {\bf r}_{0}+{\bf r}_{1}+{\bf r}_{2}+{\bf r}_{3})/4$ is a result by considerations of symmetry. It is obvious that $|{\bf R}_{0123}|<1$. The explicit form of $c_{\sigma\sigma\sigma\sigma, I}$,  $c_{\sigma\sigma\sigma\sigma, \epsilon}$  can be obtained from the calculation of 6-point correlation, say $\langle \sigma_0\sigma_1\sigma_2\sigma_3 \sigma_k \sigma_{k'}\rangle _0$, where $\sigma_k,\sigma_{k'}$ are nearest neighboring spins and far away from the defect, i.e. $|{\bf r}_k|,| {\bf r}_{k'}| \gg 1$ \cite{francesco}.  However we don't need the explicit form of these coefficients here.

Substituting it into Eq. (\ref{eq:delta-u}), we get a energy correlator $\left\langle \epsilon({\bf R}_{0123}))\epsilon({\bf R}_{ll'})\right\rangle _0\sim 1/r_l^2$, where $l,l'$ are over the whole system. All the four spins terms in Eq. (\ref{eq:expansion}) can be dealt similarly. Therefore all the terms in $\delta U$ are proportional to the integration of the energy correlator.

To be more explicit, we see the scaling further. At the critical point $\left\langle \epsilon(0)\epsilon({\bf r})\right\rangle _0\propto
1/r^2$, then the summation over the system leads to
\begin{equation}
\delta U \sim \sum_{<ll'>} \frac{1}{r^2_{l}}\propto \log N
\label{eq:logn}
\end{equation}
if the system size is $N$. Near the critical point,  $\left\langle \epsilon(0)\epsilon({\bf r}\right\rangle _0\propto e^{-r/\xi_{\pm}}/r^2$. If $N \gg \xi$, the summation over the system gives rise to
\begin{equation}
\delta U \sim \sum_{<ll'>} \frac{e^{r_l/\xi_{\pm}}}{r^2_{l}}\propto \log \xi_{\pm} \sim -\log |t|
\label{eq:delta-u-ising}
\end{equation}
This is just the behaviour the specific heat of a pure system. Then we have
\begin{equation}
\delta U \propto c
\end{equation}

As we can see the influence of the defect on the internal energy is an integration of energy correlator, which gives rise the fluctuation of energy. The physical nature of specific heat is just the fluctuation of energy. Due to the long range correlation, the defect changes the energy density across the system. The critical exponent of the internal energy variation is the same as that of the specific heat of the pure system.

Since $\delta C=\frac{\partial \delta U}{\partial \beta}$, it has
\begin{equation}
\delta C \sim \frac{\partial c}{\partial t}\propto |t|^{-1}.
\label{eq:delta-c-ising}
\end{equation}
We simply get the heat capacity variation. Using finite size scaling \cite{privman}, at the critical point, we get
\begin{equation}
\delta C \propto N.
\label{eq:n}
\end{equation}

\subsection{The general scaling argument for an isotropic system with a defect}

Consider a general model with short range interaction in d-dimension. 
\begin{equation}
H=\sum_{<ij>} J \sigma_i \sigma_j
\end{equation}
The basic operators, are the spins, for example, $\sigma_i=\pm 1$ in Ising model, $\sigma_i=(\sigma_{ix},\sigma_{iy}), \sigma_{ix}^2+\sigma_{iy}^2=1$ in XY model, etc. 

The defect will break the bonds which connect to the defect. Generally we have
\begin{equation}
e^{-\beta \delta H}=1+\delta H+\frac{1}{2} (\delta H)^2+\cdots=\sum_a b_a S_a
\end{equation}
where $S_a$ are the all possible products of $\sigma_0\sigma_i$, $i=1,2,3,4$ are the four nearest neighbours of the defect, and $b_a$ are the coefficients.  
According to the OPE, these products of basic operators can be expanded as a linear combinations of scaling operators. That is to say $S_a$ can be expanded as
\begin{equation}
S_a=\sum_k c_{ak}\phi_k({\bf r}_a)
\label{eq:ex-2}
\end{equation}
where $\phi_k$ are the scaling operators and $c_{ak}$ are the OPE coefficients . The position ${\bf r}_a$ can be determined according to OPE in Eq. (\ref{eq:ope}) step by step. Because the position ${\bf r}_i$ of the basic operators $\sigma_i$ at the defect and its nearest neighbours satisfy $|{\bf r}_i|\le 1$ if we set the lattice constant be the unit length, it should have $|{\bf r}_a|\le 1$. 

For the scaling operator $\phi_k$, the scaling dimension is $x_k=d-y_k$, where $d$ is the spatial dimension and  $y_k$ is eigenvalue of $u_k$ in the RG. At the critical point the correlation between two scaling operators $\phi_k(0),\phi_l({\bf r})$ decays as
\begin{equation}
\left\langle \phi_k(0)\phi_l({\bf r})\right\rangle _0\sim  \frac{1}{r^{x_k+x_l}}
\end{equation}
in the limit of $r \rightarrow \infty$. For a critical system, the most relevant operator are magnetic operator $s({\bf r})$ and thermal operators (or energy operators) $\epsilon({\bf r})$. Their eigenvalues are $y_h$ and $y_t$ respectively and $y_h,y_t>0$. The other scaling operators' eigenvalues are negative and hence irrelevant.

In the expansion Eq. (\ref{eq:ex-2}), the magnetic operator $s({\bf r})$ is prohibited because of symmetry. We may classify the scaling operators as being even or odd under the symmetry $\sigma_i \rightarrow -\sigma_i$. In the expansion Eq. (\ref{eq:ex-2}), the scaling operators $\phi_k({\bf r}_a)$ must be even. Besides the constant term, the first order term is the thermal operator $\epsilon({\bf r}_a)$. Then, for the internal energy variation defined in Eq. (\ref{eq:delta-u}), the leading term is a summation of energy correlator between the defect and other bonds. The correlator of thermal operators near the critical point behaves  
\begin{equation}
\left\langle \epsilon (0) \epsilon ({\bf r})\right\rangle _0\sim \frac{e^{-r/\xi_{\pm}}}{r^{2(d-y_t)}}=\frac{e^{-r/\xi_{\pm}}}{r^{2(d-1/\nu)}}.
\end{equation}
where $\nu=1/y_t$ is the correlation length exponent and $\xi_{\pm} \sim |t|^{-\nu}$. The correlators between thermal operator and other even scaling operators decays faster since the scaling dimension of other even operators are larger than thermal operator's.

Therefore at the critical point for a finite system with size $N$ and $N \gg 1$ , the internal energy variation is given by
\begin{equation}
\delta U\sim \int_1^N \frac{r^{d-1}dr}{r^{2d-2/\nu}}\sim A+N^{d-2/\nu}=A+N^{\alpha/\nu} 
\end{equation}
where $\alpha=2-d\nu$ is the specific heat exponent, $A$ is a constant, $\alpha=0$ is the logarithmic case.  If $\nu \ge 0$, the constant term $A$ is negligible, and $\delta U$ is divergent as $N \rightarrow \infty$. Otherwise $A$ is the leading term and $\delta U$ converges to $A$ as $N \rightarrow \infty$. For a finite system the specific heat at the critical point just scales as $c(L)\sim A+L^{\alpha/\nu}$.

For away from the critical point and $N\gg \xi_{\pm}$, we have
\begin{equation}
\delta U\sim \int_1^\xi \frac{r^{d-1}dr}{r^{2d-2/\nu}}\sim
A'+ \xi_{\pm}^{d-2/\nu}\sim A''+|t|^{-\alpha}
\label{eq:u-scaling}
\end{equation}
Similarly if $\nu \ge 0$, the constant term $A''$ is negligible, and $\delta U$ is divergent as $|t| \rightarrow 0$. Otherwise $A''$ is the leading term and $\delta U$ converges to $A''$ as $|t| \rightarrow 0$. In this case $\delta U$ has a cusp. As we know for a pure system the specific heat scales as $c \sim A''+ |t|^{-\alpha}$ near the critical point. Therefore we have $\delta U \sim c$. 

Since $\delta C= \frac{\partial \delta U}{\partial \beta}$, it has
\begin{equation}
\delta C\sim |t|^{-\alpha -1}.
\label{eq:c-scaling}
\end{equation}
For a finite system at the critical point \cite{privman}
\begin{equation}
 \delta C\sim N^{(\alpha+1)/\nu}.
 \label{eq:c-scaling-n}
\end{equation}

The above argument is quite general except we assumed that the system is isotropic, i.e. the correlator only depends on the separation between two points. Thus conclusions in Eq. (\ref{eq:u-scaling}) and (\ref{eq:c-scaling}) are valid for common isotropic systems.

\section{Numerical solution of the two dimensional Ising model with a defect}

To test the above conclusion, we calculate the internal energy variation and specific heat variation of two dimensional Ising model with a defect numerically.
In the lattice shown in Fig. 1, the boundary condition is open, i.e., we  have four edges and four corners. For this kind of geometry, BPA is very efficient \cite{loh1,loh2,wu1,wu2,wu3}. The sketch of BPA for this problem is also shown in Fig.1. To keep the defect be the centre of lattice, we set the size be odd and the aspect ratios are also odd. In this way we can avoid the trouble stemming from the asymmetry.

BPA is very accurate and can be carried on lattices with extremely large sizes, say $2000\times 2000$.  With it, we have verified the CFT predictions on the corner free energy with free boundary condition \cite{cardy2,kleban}. The result on the central charge of Ising model  agrees with CFT in the precision of $10^{-10}$ \cite{wu1}. We also recover the aspect ratio dependence of the corner free energy in CFT theory accurately \cite{kleban,wu1}. The corner free energy with fixed boundary and mixed boundary condition has been studied by CFT recently \cite{imamura,bondesan,roberto}. With an extended BPA \cite{wu4}, we verified these CFT prediction in the accuracy $10^{-16}$ \cite{wu5}.

In the algorithms, the transformations preserve these quantities during every step. The accuracy is only limited  by the machine's accuracy. With these algorithms, we can calculate the free energy, internal energy and specific heat with the same accuracy. As discussed in reference \cite{loh2}, the time of calculation is proportional to $L^2 \times M$ if the lattice size is $M\times L$. Then the accumulation of round-off error is proportional to $L\sqrt{M}$. Therefore the accuracy can reach $10^{-28}-10^{-29}$ for the lattice with a very large size $1000\times 1000$ if all the variables are assigned in the quadruple precision format, in which the machine accuracy is $10^{-33}$.  This has been shown in \cite{wu3}.

We apply the BPA to calculate the free energy and internal energy directly. For the heat capacity, we adapt a difference approximation. In fact, the BPA for the heat capacity on the usual lattice has been developed \cite{wu3}. However, for the lattice with a defect, we meet an unknown problem in the BPA for the heat capacity. Therefore, we calculate the internal energy $U(\beta+\Delta \beta)$ and $U(\beta-\Delta \beta)$ and take $[U(\beta+\Delta \beta)-U(\beta-\Delta \beta)]/(2\Delta \beta)$ as the approximate value of $\partial U /\partial \beta$. In the numerical calculation, we set $\Delta \beta =10^{-11}$ and set all variables in quadruple precision format. Because the result of BPA for a lattice with $1000\times 1000$ can reach $10^{-29}$,  such a small difference $\Delta \beta =10^{-11}$ does not cause serious instability. Then we can get the specific heat variation in a very high accuracy.

\subsection{At the critical point}

At first, we calculate the variations at the critical point $\beta=\beta_C=\frac{1}{2}\ln(1+\sqrt{2})$. To investigate the geometrical effect, we study the rectangle with size $M\times N$ with the aspect ratios $\rho=M/N=1,3,5,7,11$. As mentioned above to keep the defect be the centre of lattice, we set the size be odd and the aspect ratios are odd. For $\rho=1,3$, we calculate $65$ data points with $31\le N \le 1021$. For $\rho=5,7,11$, the number of data points are $63,53,45$ respectively and the range of size $31\le N \le 981,781,621$ respectively.  We fit the variations within finite size scaling.

 \begin{table}
\caption{ The the fitted $A_0$ in Eq. (\ref{eq:fex}) for the free energy variation for $\rho =1,3,5,7,11$. It has $\delta_{max}<10^{-22}$. }
\begin{tabular}{cc}

\hline
 $\rho$           &  $A_0$\\
\hline
$1$           &  ~~~~~~$1.409210310117490418345(2)$\\
$3$           &  ~~~~~~$1.409210310117490418347(1) $\\
$5$           &  ~~~~~~$1.409210310117490418347(1) $\\
$7$           &  ~~~~~~$1.409210310117490418346(2) $\\
$11$           & ~~~~~~ $1.409210310117490418344(5) $ \\
\hline
\end{tabular}
\label{tab:A_0}

\end{table}

We fit the free energy variation with
\begin{equation}
\delta F=\sum_{k=0}^{k_{max}}A_{k}N^{-k}
\label{eq:fex}
\end{equation}
In our fitting it has $k_{max}=12$. This indicates that our numerical calculation is very accurate. The first term $A_0$ gives the defect's contribution to the free energy in the thermodynamic limit $N \rightarrow \infty$. We show the first term for $\rho =1,3,5,7,11$ in Tab. I. In the thermodynamic limit, the (bulk) free energy density for a pure Ising model (without the defect) is given by $f_{bulk}=0.929695398\cdots$ at the critical point \cite{onsager}. The contribution of the defect to the free energy is about $A_0/f_{bulk}\approx 1.5$ times of the bulk free energy density. The other fitted parameters can be seen in appendix A. The high order corrections stem from the boundary and finite size effect. Only $A_0$ is boundary independent. That is to say, if someone solve this problem with periodic boundary condition, the $A_0$ should be the same and the higher order corrections may be different.

The fitting method is the Levenberg-Marquardt method \cite{press} for nonlinear fit. To characterize the accuracy of our fittings, we define the maximal deviation
\begin{equation}
\delta_{max}=Max{|y_i-y_i^{fit}|},
\label{eq:error}
\end{equation}
 where $y_i$ is the numerical data and $y_i^{fit}$ is the value given by the fitting formula. We choose the maximum of the deviations from the data to the fitted ones to represent our fitting quality. For a given fitting formula , the deviations are given by the nonlinear fitting algorithm.  We expand the the free energy, internal energy and specific heat to as high order as possible to make the $\delta_{max}$ as small as possible. In every table of the fitting parameters, we give the maximum of deviation. For example, in Tab. \ref{tab:A_0}, $\delta_{max}=10^{-22}$. 

As mentioned above, the results of BPA for $F,F_0,U,U_0$ are in an accuracy of $10^{-28}-10^{-29}$. The fitting accuracies of $\delta F, \delta U$ are about $10^{-22}-10^{-23}$. They are consistent because $F,F_0,U,U_0$ are in order of $10^{6}$ for $1000\times 1000$ lattice, the first six digits of $F,F_0$ are cancelled in $\delta F$ and it is the same for $U,U_0$ and $\delta U$. This is why the maximal deviations in Table I-X are $10^{-22}-10^{-23}$. For heat capacity variation, we use $\delta C \approx [\delta U(\beta+\Delta \beta)-\delta U(\beta-\Delta \beta)]/(2\Delta \beta)$ and $\Delta \beta=10^{-11}$, then the accuracy of $\delta C$ should be in order of $10^{-11}-10^{-12}$, since the deviation of $\delta U$ is multiplied by $1/\Delta \beta$. In the fittings of $\delta C$, the maximal deviation is about $10^{-12}$ as shown in Tables XI-XV. This is also consistent.

The internal energy variation can be fitted by the formula
\begin{equation}
\delta U=\sum_{k=0}^{k_{max}}  \frac{B_{1k}\ln N+B_{0k}}{N^{-k}}
\label{eq:uex}
\end{equation}
In our fitting, $k_{max}$ is set to be $8$.  The leading term diverges logarithmically with the system size. This leading term is obtained from Eq. (\ref{eq:logn}).  The coefficient of leading term $B_{10}$ is shown in Tab. II. As we can see the leading term is geometry independent, i.e. $B_{10}$ are the same for $\rho=1,3,5,7,11$ in the error range $<10^{-15}$.  The higher order terms can be seen in the appendix B. $B_{10}$ should be independent of the boundary condition. Therefore, if someone solve this problem with periodic boundary condition or other boundary condition, the $B_{10}$ should be the same and the higher order corrections may be different. 

\begin{table}
\caption{ The fitted $B_{10}$ in Eq. (\ref{eq:uex}) the internal energy variation for $\rho =1,3,5,7,11$. It has $\delta_{max}<10^{-22}$. }
\begin{tabular}{cc}

\hline
 $\rho$           &  $B_{10}$\\
\hline
$1$           &  ~~~~~$1.7560009405073355(2)$\\
$3$           &  ~~~~~~~$1.75600094050733606(1) $\\
$5$           &  ~~~~~~~$1.75600094050733605(1) $\\
$7$           &  ~~~~~$1.7560009405073360(2) $\\
$11$          &  ~~~~$1.756000940507336(5) $ \\
\hline
\end{tabular}
\label{tab:B_0}

\end{table}

\begin{table}
\caption{ The fiited  $C_{-1}$ and  $C_{20}$  in Eq. (\ref{eq:cex}) for the specific heat variation for $\rho =1,3,5,7,11$. It has $\delta_{max}<10^{-12}$. }
\begin{tabular}{ccc}

\hline
 $\rho$           &  $C_{-1}$ &  $C_{20}$\\
\hline
$1$            & ~~~~$0.370131893446(5)$ &  ~~$0.598836(1)$\\
$3$           &  ~~~~$0.473858336337(2) $ &  ~~~~~$0.59883789(1)$\\
$5$           &  ~~~~$0.475120770039(2) $ &  ~~~~~$0.59883783(2)$\\
$7$           &  ~~~~$0.475126998263(7) $ &  ~~~~~$0.59883752(4)$\\
$11$          &  ~~~~$0.475127020038(2) $ &  $0.59884(1)$ \\
\hline
\end{tabular}
\label{tab:C_0}

\end{table}

We fit the heat capacity variation in the following way
\begin{equation}
\delta C=C_{-1}N+\sum_{k=0}^{k_{max}}  \frac{C_{2k}\ln^2 N+C_{1k}\ln N+C_{0k}}{N^{-k}}
\label{eq:cex}
\end{equation}
In our fitting, $k_{max}$ is set to be $6$. The leading term $C_{-1}$ and $C_{20}$ are shown in Tab. III. The heat capacity variation shows a more dramatic effect. It diverges linearly with the size! This leading divergent term is obtained from Eq. (\ref{eq:n}) in the last section. The coefficients of this linear term depend on the aspect ratio $\rho$, i.e. they are geometry dependent. The next leading term is $C_{20}\ln ^2 N$.  Contrary to the first leading term, this term $C_{20}$ is independent of the aspect ratio, at least in the error range. The other fitted parameters can be seen in the appendix C. This fitting formula, especially for the term $\ln ^2 N$, is a bit strange. We have tried many different formulas to fit the data. Only with this formula, we can fit the data with the smallest deviation. 

One may note that the geometry in our consideration as shown in Fig. 1 has open boundary: open edges and sharp corners. These boundaries will produce edge terms and corner terms \cite{privman, cardy, wu1}. However, in our situation, the boundary effects are cancelled exactly since both $F,U,C$ and $F_0,U_0,C_0$ contain the same boundary terms. 
We have expanded $F, F_0$ separately and found that their edge terms are the same and so are their corner terms. It is the same for $U,U_0$ and $C,C_0$.
Hence in $\delta F, \delta U , \delta C$ there are no boundary terms.

From the fitting at the critical point, we get  $\delta U \sim \ln N$ and $\delta C \sim N$. This verifies the conclusions given in Eq. (\ref{eq:logn}) and (\ref{eq:n}).

\subsection{Temperature dependence of the variations}

\begin{figure}
\includegraphics[width=0.5\textwidth]{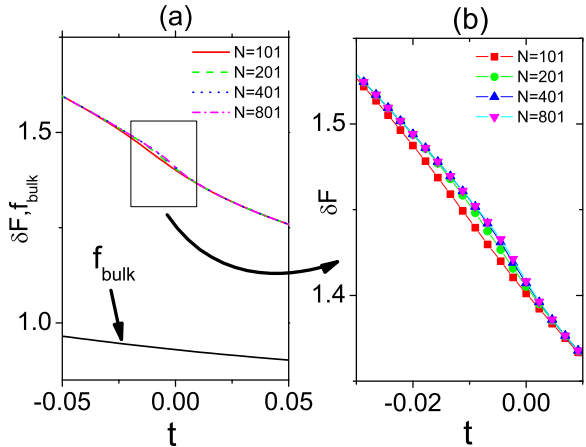}
\caption{ (a) The free energy variation in the critical regime. (b) The finite size effect on the free energy variations.  }
\end{figure}

To study the finite size scaling, we calculate the variations for square lattices with size $N=101,201,401,801$. Fig. 2 shows the variation of the free energy in the critical regime for square lattice with these sizes . In order to see the ratio between $\delta F$ and the bulk free energy density $f_{bulk}$, we also show the bulk free energy density \cite{onsager} with black line in Fig. 2a. Near the critical point, the ratio $\delta F/f_{bulk}$ is about $1.5$ for $N>100$. As we can see these results almost coincide except for very small $t$ because the lattice sizes are very large. Fig. 2b shows the convergence of the results of different size at very small $t$.

It is should be emphasized that we use directly the reduced temperature $t=(T-T_c)/T_c$, where $T_c=2/\ln(1+\sqrt{2})=2.2691853 \cdots$ is the infinite volume $T_c$.  Because the lattice sizes $N=101,201,401,801$ are large enough so that the finite-size critical temeprature is very well described by the infinite volume $T_c$.

\begin{figure}
\includegraphics[width=0.5\textwidth]{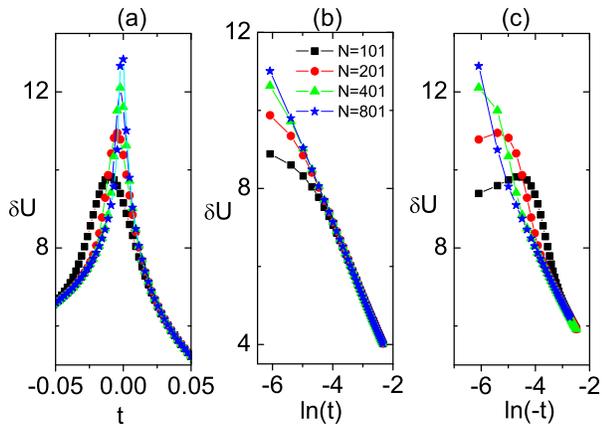}
\caption{ (a) The internal energy variation in the critical regime. (b) The logarithmic dependence of the reduced temperature for $t>0$. (c)The logarithmic dependence of the reduced temperature for $t<0$. The legend in (a) and (c) is the same as that in (b). }
\end{figure}

\begin{figure}
\includegraphics[width=0.5\textwidth]{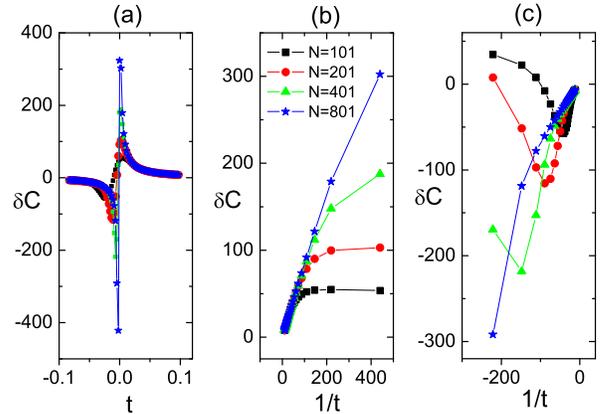}
\caption{ (a) The heat capacity variation in the critical regime. (b) The inverse dependence of the reduced temperature for $t>0$. (c)The inverse dependence of the reduced temperature for $t<0$. The legend in (a) and (c) is the same as that in (b).   }
\end{figure}

To verify Eq. (\ref{eq:delta-u-ising}) and (\ref{eq:delta-c-ising}) with the the finite size scaling, we also study the dependence of the variations on the temperature.
Fig. 3 shows the internal energy variation for the lattices with size $N=101,201,401,801$.  Fig. 3a shows the logarithmic divergence. In order to show this divergence more clearly, we show the part for $t>0$ and $t<0$ in Fig. 3b and 3c respectively. The linear dependence of the variation on the $\ln |t|$ can be seen. As the size increases, the linear region grows. Fitting the linear parts for $N=801$, we get
\begin{equation}
\delta U \approx B' \ln |t|
\end{equation}
where $B'=1.888(4)$ for $0.003<t<0.1$ and $B'=1.68(5)$ for $0.003<|t|<0.1$ and $t<0$. This agrees with the expansion at the critical point, where $B_{10}=1.75 \cdots$.

Fig. 4 shows the specific heat variation for the lattices with size $N=101,201,401,801$.  Fig. 4a shows the global feature in the critical region. The diverging trend as $t$ approaches $0$ can be clearly seen. In order to show this divergence more clearly, we show the part for $t>0$ and $t<0$ in Fig. 4b and 4c respectively. The linear dependence of the variation on the $|t|^{-1}$ is shown. As the size increases, the linear region grows.
Fitting the linear parts for $N=800$, we get
\begin{equation}
\delta C \approx C' |t|^{-1}
\end{equation}
where $C'=1.8$ for $t>0$ and $C'=1.5$ for $t<0$.

As we know, near the critical point it has $c\sim \ln |t|$ and $\frac{\partial c}{\partial t} \sim |t|^{-1}$ for the two dimensional Ising model \cite{onsager}. The finite size scaling near the critical point verifies the conclusions given in Eq. (\ref{eq:delta-u-ising}) and (\ref{eq:delta-c-ising}) again.

\section{Summary and acknowledgement}

We show  general properties for the critical systems with a single defect. The internal energy variation is proportional to the specific heat of the pure system and the heat capacity variation is proportional to the temperature derivative of the specific heat. The numerical calculation for two dimensional Ising model with a site defect verifies these  properties.
We present a theoretical approach to describe a very small consequence on thermodynamic quantities of a defect and their FSS relations although the numerical variations are very small

The effect of a defect is relatively small if the system is macroscopic. However if the system  size is small, the defect's effect becomes remarkable.  Besides, there are some advantages in measuring $\delta U, \delta C$.

1. To find the defect'e effect, the relative accuracy needs not to be so high. For example, consider a $100 \times 100$ lattice (made with cold atom technique).
For a non-critical system, if one atom is removed, the effect can be discovered only if the relative accuracy reaches 1/10000. 
However, if the system is Ising-like and at the critical point, the effect can be discovered if the relative accuracy of heat capacity reaches about $ 1/570$.
This is because the total heat capacity is about $0.45 \times 10000 \times \ln100$ and the variation of the heat capacity is $0.37\times 100$ according to our numerical result.

2. Another remarkable feature of $\delta C$ is that it changes sign as the temperature crosses the critical temperature.

3. The boundary effects are cancelled in $\delta U$ because $U,U_0$ have the same boundary effects. It is the same for $\delta C$.

Modern nano technology and technology of cold atom  makes it possible to test these predictions. We are expecting the experiments on this topic.

The authors thank Joseph Indekeu, Zehui Deng, Adoardo Lauria, Fedorico Galli for useful discussions. 

\newpage
\appendix

\section{Tables for fitted parameters in Eq. (38)}

Table IV-VIII show the fitted parameters in Eq. (38) for the free energy variation with $\rho=1,3,5,7,11$.

\begin{table}
 \caption{ The fitted parameters in Eq. (38) for the free energy variation with $\rho=1$. It has $\delta_{max}<10^{-23}$. }
\begin{tabular}{crr}

\hline
            &  $A_i$~~~~~~~~~~~~~~~~~~~~~~~~~~~~~      &  $\Delta A_i$ ~~~~\\
\hline

$A_0   $  &    0.1409210310117490418347951D+01&     0.2D$-$21 \\
$A_1   $  &   -0.8139392192780112125175834D+00&     0.5D$-$18 \\
$A_2   $  &    0.9067904677646148699557989D+00&     0.5D$-$15 \\
$A_3   $  &   -0.1951243710139618217363658D+01&     0.3D$-$12 \\
$A_4   $  &    0.5044255710564309572237116D+01&     0.9D$-$10 \\
$A_5   $  &   -0.1097266684518414528324830D+02&     0.2D$-$07 \\
$A_6   $  &    0.2276143211351712932688598D+02&     0.3D$-$05 \\
$A_7   $  &   -0.5145819510473408538160129D+02&     0.3D$-$03 \\
$A_8   $  &    0.2111822173858678998575746D+03&     0.2D$-$01 \\
$A_9   $  &   -0.8914083610684714743102263D+03&     0.8D+00 \\
$A_{10}$  &    0.3184490995226739488807939D+04&     0.2D+02 \\
$A_{11}$  &   -0.9963717158837611501676144D+04&     0.4D+03 \\
$A_{12}$  &    0.3453273003867273296055979D+05&     0.3D+04 \\

\hline
\end{tabular}
\label{tab:f1}
\end{table}

\begin{table}
 \caption{ The fitted parameters in Eq. (38) for the free energy variation with $\rho=3$. It has $\delta_{max}<10^{-23}$. }
\begin{tabular}{crr}

\hline
            &  $A_i$~~~~~~~~~~~~~~~~~~~~~~~~~~~~~      &  $\Delta A_i$ ~~~~\\
\hline

$A_0   $  &    0.1409210310117490418348039D+01&     0.2D$-$21 \\
$A_1   $  &   -0.6898025698420019143116894D+00&     0.5D$-$18 \\
$A_2   $  &    0.7246887439974793787485822D+00&     0.5D$-$15 \\
$A_3   $  &   -0.1163435600677590595858805D+01&     0.3D$-$12 \\
$A_4   $  &    0.2125184926326692099458813D+01&     0.9D$-$10 \\
$A_5   $  &   -0.4104674530813393591114680D+01&     0.2D$-$07 \\
$A_6   $  &    0.1019038004291499836646201D+02&     0.3D$-$05 \\
$A_7   $  &   -0.2695588248794192948700816D+02&     0.3D$-$03 \\
$A_8   $  &    0.9263301940770832606549970D+02&     0.2D$-$01 \\
$A_9   $  &   -0.3477772253050403996019055D+03&     0.9D+00 \\
$A_{10}$  &    0.1584171365328676077346061D+04&     0.2D+02 \\
$A_{11}$  &   -0.6764747691605833782586485D+04&     0.4D+03 \\
$A_{12}$  &    0.2185652903261572286215693D+05&     0.3D+04 \\

\hline
\end{tabular}
\label{tab:f3}
\end{table}

\begin{table}
 \caption{ The fitted parameters in Eq. (38) for the free energy variation with $\rho=5$. It has $\delta_{max}<10^{-23}$. }
\begin{tabular}{crr}

\hline
            &  $A_i$~~~~~~~~~~~~~~~~~~~~~~~~~~~~~      &  $\Delta A_i$ ~~~~\\
\hline

$A_0   $  &    0.1409210310117490418348008D+01&     0.2D$-$21 \\
$A_1   $  &   -0.6895803724829525587022584D+00&     0.5D$-$18 \\
$A_2   $  &    0.7253638089447714319588130D+00&     0.5D$-$15 \\
$A_3   $  &   -0.1160074030138000567633846D+01&     0.3D$-$12 \\
$A_4   $  &    0.2124723799806189308335581D+01&     0.1D$-$09 \\
$A_5   $  &   -0.4099892626938691331920487D+01&     0.2D$-$07 \\
$A_6   $  &    0.1015677717392225622184707D+02&     0.3D$-$05 \\
$A_7   $  &   -0.2690303802284446466680259D+02&     0.3D$-$03 \\
$A_8   $  &    0.9253410502505634858180149D+02&     0.2D$-$01 \\
$A_9   $  &   -0.3474282708150724435713793D+03&     0.9D+00 \\
$A_{10}$  &    0.1580378131891456409396060D+04&     0.3D+02 \\
$A_{11}$  &   -0.6710613069181165806211326D+04&     0.4D+03 \\
$A_{12}$  &    0.2149053023432619000704101D+05&     0.3D+04 \\

\hline
\end{tabular}
\label{tab:f5}
\end{table}

\begin{table}
 \caption{ The fitted parameters in Eq. (38) for the free energy variation with $\rho=7$. It has $\delta_{max}<10^{-23}$. }
\begin{tabular}{crr}

\hline
            &  $A_i$~~~~~~~~~~~~~~~~~~~~~~~~~~~~~      &  $\Delta A_i$ ~~~~\\
\hline

$A_0   $  &    0.1409210310117490418348016D+01&     0.5D$-$21 \\
$A_1   $  &   -0.6895799575755920995042567D+00&     0.1D$-$17 \\
$A_2   $  &    0.7253669127704436033610930D+00&     0.1D$-$14 \\
$A_3   $  &   -0.1160055610993603474765616D+01&     0.5D$-$12 \\
$A_4   $  &    0.2124777943031982706300862D+01&     0.2D$-$09 \\
$A_5   $  &   -0.4099763470545508107346941D+01&     0.3D$-$07 \\
$A_6   $  &    0.1015685122021770300170225D+02&     0.5D$-$05 \\
$A_7   $  &   -0.2690305830117266022630290D+02&     0.4D$-$03 \\
$A_8   $  &    0.9253334141723530059010477D+02&     0.3D$-$01 \\
$A_9   $  &   -0.3474247146179242454472337D+03&     0.1D+01 \\
$A_{10}$  &    0.1580295378288385220823113D+04&     0.3D+02 \\
$A_{11}$  &   -0.6709437895591080697242263D+04&     0.5D+03 \\
$A_{12}$  &    0.2148323931741508684879155D+05&     0.4D+04 \\

\hline
\end{tabular}
\label{tab:f7}
\end{table}

\begin{table}
 \caption{ The fitted parameters in Eq. (38) for the free energy variation with $\rho=11$. It has $\delta_{max}<10^{-23}$. }
\begin{tabular}{crr}

\hline
            &  $A_i$~~~~~~~~~~~~~~~~~~~~~~~~~~~~~      &  $\Delta A_i$ ~~~~\\
\hline

$A_0   $  &    0.1409210310117490418348135D+01&     0.1D$-$20 \\
$A_1   $  &   -0.6895799567993295580763121D+00&     0.2D$-$17 \\
$A_2   $  &    0.7253669220327549364870848D+00&     0.2D$-$14 \\
$A_3   $  &   -0.1160055538390475552667760D+01&     0.9D$-$12 \\
$A_4   $  &    0.2124778303511495146132637D+01&     0.3D$-$09 \\
$A_5   $  &   -0.4099762126894845643279945D+01&     0.5D$-$07 \\
$A_6   $  &    0.1015685523707872014193952D+02&     0.7D$-$05 \\
$A_7   $  &   -0.2690310124826811330621585D+02&     0.7D$-$03 \\
$A_8   $  &    0.9253638328030327623722488D+02&     0.4D$-$01 \\
$A_9   $  &   -0.3475505564770179303424906D+03&     0.2D+01 \\
$A_{10}$  &    0.1583637648017210503021256D+04&     0.5D+02 \\
$A_{11}$  &   -0.6760622483924301563712628D+04&     0.7D+03 \\
$A_{12}$  &    0.2182556528199001879566179D+05&     0.5D+04 \\

\hline
\end{tabular}
\label{tab:f11}
\end{table}

\newpage
\section{Tables for Eq. (40)}

Table IX-XIII show the fitted parameters in Eq. (40) for the internal energy variation with $\rho=1,3,5,7,11$.

\begin{table}
 \caption{ The fitted parameters in Eq. (40) for the internal energy variation with $\rho=1$. It has $\delta_{max}<10^{-23}$. }
\begin{tabular}{crr}

\hline
            &  $B_i$~~~~~~~~~~~~~~~~~~~~~~~~~~~~~      &  $\Delta B_i$ ~~~~\\
\hline

$B_{10}$  &    0.1756000940507335790614587D+01&     0.5D$-$16 \\
$B_{00}$  &    0.1108789097300574644698700D+01&     0.5D$-$15 \\
$B_{11}$  &   -0.1429278034573753818480900D+01&     0.7D$-$12 \\
$B_{01}$  &    0.6499233012947408899114641D+00&     0.6D$-$11 \\
$B_{12}$  &    0.2173997621082043760717310D+01&     0.2D$-$08 \\
$B_{02}$  &   -0.2578697277246965898741373D+01&     0.1D$-$07 \\
$B_{13}$  &   -0.4426808373674998573819779D+01&     0.1D$-$05 \\
$B_{03}$  &    0.5993117779650739766451307D+01&     0.8D$-$05 \\
$B_{14}$  &    0.1153234993606497236496378D+02&     0.4D$-$03 \\
$B_{04}$  &   -0.1059305245056648116716548D+02&     0.2D$-$02 \\
$B_{15}$  &   -0.2839047458341355573904127D+02&     0.4D$-$01 \\
$B_{05}$  &    0.2106973174766726014976398D+02&     0.2D+00 \\
$B_{16}$  &    0.2524070175102941639006634D+02&     0.1D+01 \\
$B_{06}$  &    0.9182572691186146023275088D+02&     0.5D+01 \\
$B_{17}$  &   -0.7718165760748037716462095D+03&     0.2D+02 \\
$B_{07}$  &    0.1508727615365346140113329D+04&     0.3D+02 \\
$B_{18}$  &   -0.1443142955304667511037682D+04&     0.4D+02 \\
$B_{08}$  &   -0.1211096405529147735468632D+04&     0.2D+02 \\

\hline
\end{tabular}
\label{tab:u1}
\end{table}

\begin{table}
 \caption{ The fitted parameters in Eq. (40) for the internal energy variation with $\rho=3$. It has $\delta_{max}<10^{-23}$. }
\begin{tabular}{crr}

\hline
            &  $B_i$~~~~~~~~~~~~~~~~~~~~~~~~~~~~~      &  $\Delta B_i$ ~~~~\\
\hline

$B_{10}$  &    0.1756000940507336069227175D+01&     0.5D$-$16 \\
$B_{00}$  &    0.1442635599708725401635079D+01&     0.5D$-$15 \\
$B_{11}$  &   -0.1211293961408149151708678D+01&     0.7D$-$12 \\
$B_{01}$  &    0.5043655667225743104254069D+00&     0.6D$-$11 \\
$B_{12}$  &    0.1690330956193629515035421D+01&     0.2D$-$08 \\
$B_{02}$  &   -0.1182276963774323003930241D+01&     0.1D$-$07 \\
$B_{13}$  &   -0.2740849022641571930683537D+01&     0.1D$-$05 \\
$B_{03}$  &    0.1404395814863511902613290D+01&     0.8D$-$05 \\
$B_{14}$  &    0.5181119451810307304256970D+01&     0.4D$-$03 \\
$B_{04}$  &   -0.2195324888537194694893548D+01&     0.2D$-$02 \\
$B_{15}$  &   -0.1063018243120927121377483D+02&     0.4D$-$01 \\
$B_{05}$  &    0.6380322280328616256054328D+01&     0.2D+00 \\
$B_{16}$  &    0.1680697493821303414490378D+02&     0.1D+01 \\
$B_{06}$  &    0.1559103790846971099277808D+02&     0.5D+01 \\
$B_{17}$  &   -0.1981485764840919333269028D+03&     0.2D+02 \\
$B_{07}$  &    0.3337347563283782000262181D+03&     0.3D+02 \\
$B_{18}$  &   -0.2418971209888612849943305D+03&     0.4D+02 \\
$B_{08}$  &   -0.2713954613779996002250673D+03&     0.2D+02 \\

\hline
\end{tabular}
\label{tab:u3}
\end{table}

\begin{table}
 \caption{ The fitted parameters in Eq. (40) for the internal energy variation with $\rho=5$. It has $\delta_{max}<10^{-22}$. }
\begin{tabular}{crr}

\hline
            &  $B_i$~~~~~~~~~~~~~~~~~~~~~~~~~~~~~      &  $\Delta B_i$ ~~~~\\
\hline

$B_{10}$  &    0.1756000940507336058650080D+01&     0.6D$-$16 \\
$B_{00}$  &    0.1443994407066697150440489D+01&     0.7D$-$15 \\
$B_{11}$  &   -0.1210903782636823168187756D+01&     0.9D$-$12 \\
$B_{01}$  &    0.5091515461841247195439272D+00&     0.7D$-$11 \\
$B_{12}$  &    0.1691247267561949045430528D+01&     0.2D$-$08 \\
$B_{02}$  &   -0.1161748603765854362514163D+01&     0.2D$-$07 \\
$B_{13}$  &   -0.2735654880042078251899909D+01&     0.2D$-$05 \\
$B_{03}$  &    0.1410024855708987543124567D+01&     0.1D$-$04 \\
$B_{14}$  &    0.5176567746969383580895496D+01&     0.4D$-$03 \\
$B_{04}$  &   -0.2179708739966420878693034D+01&     0.2D$-$02 \\
$B_{15}$  &   -0.1062524322352045950670683D+02&     0.4D$-$01 \\
$B_{05}$  &    0.6260914202077016743729399D+01&     0.2D+00 \\
$B_{16}$  &    0.1649388105790230667518388D+02&     0.2D+01 \\
$B_{06}$  &    0.1651801479893153943505287D+02&     0.5D+01 \\
$B_{17}$  &   -0.2007530400126966705185987D+03&     0.2D+02 \\
$B_{07}$  &    0.3381149739344958924087903D+03&     0.3D+02 \\
$B_{18}$  &   -0.2476407744793499744789163D+03&     0.4D+02 \\
$B_{08}$  &    0.2741340607339576860372312D+03&     0.2D+02 \\

\hline
\end{tabular}
\label{tab:u5}
\end{table}

\begin{table}
 \caption{ The fitted parameters in Eq. (40) for the internal energy variation with $\rho=7$. It has $\delta_{max}<10^{-22}$. }
\begin{tabular}{crr}

\hline
            &  $B_i$~~~~~~~~~~~~~~~~~~~~~~~~~~~~~      &  $\Delta B_i$ ~~~~\\
\hline

$B_{10}$  &    0.1756000940507336016832686D+01&     0.3D$-$15 \\
$B_{00}$  &    0.1443998353193895611366441D+01&     0.3D$-$14 \\
$B_{11}$  &   -0.1210903054059646131733464D+01&     0.3D$-$11 \\
$B_{01}$  &    0.5091822951868815096639522D+00&     0.3D$-$10 \\
$B_{12}$  &    0.1691252214235173933684408D+01&     0.7D$-$08 \\
$B_{02}$  &   -0.1161571472636836367442673D+01&     0.5D$-$07 \\
$B_{13}$  &   -0.2735626953648101518756340D+01&     0.5D$-$05 \\
$B_{03}$  &    0.1410560164873363940494722D+01&     0.3D$-$04 \\
$B_{14}$  &    0.5176421862069352651115395D+01&     0.1D$-$02 \\
$B_{04}$  &   -0.2177402993806319108001286D+01&     0.6D$-$02 \\
$B_{15}$  &   -0.1064664254167213280768083D+02&     0.1D+00 \\
$B_{05}$  &    0.6350090315565562597904054D+01&     0.5D+00 \\
$B_{16}$  &    0.1571933450600032936027599D+02&     0.4D+01 \\
$B_{06}$  &    0.1886927025564315622330318D+02&     0.1D+02 \\
$B_{17}$  &   -0.2092958846293486710359762D+03&     0.5D+02 \\
$B_{07}$  &    0.3525515468089597601811773D+03&     0.8D+02 \\
$B_{18}$  &   -0.2646422214482911819703072D+03&     0.9D+02 \\
$B_{08}$  &   -0.2846318864144246813607891D+03&     0.6D+02 \\

\hline
\end{tabular}
\label{tab:u7}
\end{table}

\begin{table}
 \caption{ The fitted parameters in Eq. (40) for the internal energy variation with $\rho=11$. It has $\delta_{max}<10^{-22}$. }
\begin{tabular}{crr}

\hline
            &  $B_i$~~~~~~~~~~~~~~~~~~~~~~~~~~~~~      &  $\Delta B_i$ ~~~~\\
\hline

$B_{10}$  &    0.1756000940507336137644741D+01&     0.1D$-$14 \\
$B_{00}$  &    0.1443998363217383095050024D+01&     0.1D$-$13 \\
$B_{11}$  &   -0.1210903052695266140292354D+01&     0.1D$-$10 \\
$B_{01}$  &    0.5091824170970866970794450D+00&     0.1D$-$09 \\
$B_{12}$  &    0.1691252231953949058566169D+01&     0.3D$-$07 \\
$B_{02}$  &   -0.1161570544019369750197256D+01&     0.2D$-$06 \\
$B_{13}$  &   -0.2735625429719691217440854D+01&     0.2D$-$04 \\
$B_{03}$  &    0.1410556584368756569688147D+01&     0.1D$-$03 \\
$B_{14}$  &    0.5176729371229483185759416D+01&     0.4D$-$02 \\
$B_{04}$  &   -0.2178894119812718899344679D+01&     0.2D$-$01 \\
$B_{15}$  &   -0.1062055808299885422980119D+02&     0.4D+00 \\
$B_{05}$  &    0.6246745084990453792973751D+01&     0.2D+01 \\
$B_{16}$  &    0.1654368780944730319389318D+02&     0.1D+02 \\
$B_{06}$  &    0.1646591910591960103095858D+02&     0.4D+02 \\
$B_{17}$  &   -0.2011852559895122908091908D+03&     0.1D+03 \\
$B_{07}$  &    0.3397153318119339570554453D+03&     0.2D+03 \\
$B_{18}$  &   -0.2500642654428872245075102D+03&     0.3D+03 \\
$B_{08}$  &   -0.2742309124676317551292031D+03&     0.2D+03 \\

\hline
\end{tabular}
\label{tab:u11}
\end{table}

\newpage
\section{Tables for Eq. (41)}

Table XIV-XVIII show the fitted parameters in Eq. (41) for the heat capacity variation with $\rho=1,3,5,7,11$.

\begin{table}
 \caption{ The fitted parameters in Eq. (41) for the heat capacity variation with $ \rho=1$. It has $\delta_{max}<10^{-12}$. }
\begin{tabular}{crr}

\hline
            &  $C_i$~~~~~~~~~~~~~~~~~~~~~~~~~~~~~      &  $\Delta C_i$ ~~~~\\
\hline

$C_{-1}$  &    0.3701318932736015169025661D+00&     0.2D$-$11 \\
$C_{20}$  &    0.5988365596873042415899176D+00&     0.1D$-$07 \\
$C_{10}$  &    0.1362181075220199930949431D$-$01&     0.3D$-$06 \\
$C_{00}$  &    0.1564365611908562201157505D+00&     0.1D$-$05 \\
$C_{21}$  &   -0.9698724835676889360365375D+00&     0.4D$-$04 \\
$C_{11}$  &    0.3700976030094435769966437D+00&     0.4D$-$03 \\
$C_{01}$  &    0.4738066353822200246665512D$-$01&     0.2D$-$02 \\
$C_{22}$  &    0.1369062477761004233681053D+01&     0.4D$-$02 \\
$C_{12}$  &    0.1055363190577289429792387D+01&     0.3D$-$01 \\
$C_{02}$  &    0.6685542891174151658098970D+01&     0.7D$-$01 \\
$C_{23}$  &   -0.5617895569195019217213644D+01&     0.2D$-$01 \\
$C_{13}$  &    0.1968401777398099423864379D+02&     0.3D$-$01 \\
$C_{03}$  &   -0.4345729961143958935258021D+01&     0.8D$-$01 \\

\hline
\end{tabular}
\label{tab:c1}
\end{table}

\begin{table}
 \caption{ The fitted parameters in Eq. (41) for the heat capacity variation with $\rho=3$. It has $\delta_{max}<10^{-13}$. }
\begin{tabular}{crr}

\hline
            &  $C_i$~~~~~~~~~~~~~~~~~~~~~~~~~~~~~      &  $\Delta C_i$ ~~~~\\
\hline

$C_{-1}$  &    0.4738583363371669290171549D+00&     0.2D$-$10 \\
$C_{20}$  &    0.5988378922929787475482076D+00&     0.1D$-$06 \\
$C_{10}$  &    0.2412944175798466397089702D+00&     0.3D$-$05 \\
$C_{00}$  &    0.2240269722942353671837937D+00&     0.1D$-$04 \\
$C_{21}$  &   -0.8253342873928658353818042D+00&     0.4D$-$03 \\
$C_{11}$  &    0.1680913744352078951227394D+00&     0.4D$-$02 \\
$C_{01}$  &    0.1173278149931294500887596D+00&     0.2D$-$01 \\
$C_{22}$  &    0.1447584978923885940808479D+01&     0.4D$-$01 \\
$C_{12}$  &   -0.1110998938596696223025486D+01&     0.3D+00 \\
$C_{02}$  &    0.1239621370168218395659038D+01&     0.7D+00 \\
$C_{23}$  &   -0.4878300240223302096521914D+01&     0.2D+00 \\
$C_{13}$  &    0.6192388723449244329513208D+01&     0.3D+00 \\
$C_{03}$  &   -0.6071216223290772832895171D+01&     0.8D+00 \\

\hline
\end{tabular}
\label{tab:c3}
\end{table}

\begin{table}
 \caption{ The fitted parameters in Eq. (41) for the heat capacity variation with $\rho=5$. It has $\delta_{max}<10^{-12}$. }
\begin{tabular}{crr}

\hline
            &  $C_i$~~~~~~~~~~~~~~~~~~~~~~~~~~~~~      &  $\Delta C_i$ ~~~~\\
\hline

$C_{-1}$  &    0.4751207700394731046622277D+00&     0.2D$-$11 \\
$C_{20} $  &    0.5988378319485564075712419D+00&     0.2D$-$07 \\
$C_{10}$  &    0.2422223399882923582515648D+00&     0.3D$-$06 \\
$C_{00}$  &    0.2288063764922809820669021D+00&     0.2D$-$05 \\
$C_{21}$  &   -0.8249484783990579853710463D+00&     0.4D$-$04 \\
$C_{11}$  &    0.1693499100245821749277082D+00&     0.5D$-$03 \\
$C_{01}$  &    0.1401363136723088812062587D+00&     0.2D$-$02 \\
$C_{22}$  &    0.1437916959377947883645319D+01&     0.5D$-$02 \\
$C_{12}$  &   -0.1040371814683644478974431D+01&     0.3D$-$01 \\
$C_{02}$  &    0.1096885820297504552215861D+01&     0.8D$-$01 \\
$C_{23}$  &   -0.4853343844087464258713497D+01&     0.2D$-$01 \\
$C_{13}$  &    0.6288558469821367291050283D+01&     0.3D$-$01 \\
$C_{03}$  &   -0.5969528111506215066922130D+01&     0.9D$-$01 \\

\hline
\end{tabular}
\label{tab:c5}
\end{table}

\begin{table}
 \caption{ The fitted parameters in Eq. (41) for the heat capacity variation with $\rho=7$. It has $\delta_{max}<10^{-13}$. }
\begin{tabular}{crr}

\hline
            &  $C_i$~~~~~~~~~~~~~~~~~~~~~~~~~~~~~      &  $\Delta C_i$ ~~~~\\
\hline

$C_{-1}$  &    0.4751269982626733516613157D+00&     0.7D$-$10 \\
$C_{20}$  &    0.5988375176118030731401894D+00&     0.4D$-$06 \\
$C_{10}$  &    0.2422310974109595648084283D+00&     0.8D$-$05 \\
$C_{00}$  &    0.2288248568547370662554431D+00&     0.4D$-$04 \\
$C_{21}$  &   -0.8242508716769945382837724D+00&     0.1D$-$03 \\
$C_{11}$  &    0.1609297524430012514725075D+00&     0.1D$-$01 \\
$C_{01}$  &    0.1707615732717604233548380D+00&     0.4D$-$01 \\
$C_{22}$  &    0.1368698698729565558131830D+01&     0.1D+00 \\
$C_{12}$  &   -0.6082377549635815123532189D+00&     0.6D+00 \\
$C_{02}$  &   -0.4448815393554476152463779D-01&     0.2D+00 \\
$C_{23}$  &   -0.4593854244750506845249013D+01&     0.4D+00 \\
$C_{13}$  &    0.6752451697676445385979327D+01&     0.7D+00 \\
$C_{03}$  &   -0.4709263637062352133372814D+01&     0.2D+01 \\

\hline
\end{tabular}
\label{tab:c7}
\end{table}

\begin{table}
 \caption{ The fitted parameters in Eq. (41) for the heat capacity variation with $\rho=11$. It has $\delta_{max}<10^{-12}$. }
\begin{tabular}{crr}

\hline
            &  $C_i$~~~~~~~~~~~~~~~~~~~~~~~~~~~~~      &  $\Delta C_i$ ~~~~\\
\hline

$C_{-1}$  &    0.4751270200380996173992645D+00&     0.2D$-$08 \\
$C_{20}$  &    0.5988406681372792185253532D+00&     0.1D$-$04 \\
$C_{10}$  &    0.2421711148698039981400914D+00&     0.2D$-$03 \\
$C_{00}$  &    0.2291264283210204840237097D+00&     0.1D$-$02 \\
$C_{21}$  &   -0.8304126144164101519217354D+00&     0.3D$-$01 \\
$C_{11}$  &    0.2340267004018715595298064D+00&     0.3D+00 \\
$C_{01}$  &   -0.8861387725813416786723828D-01&     0.1D+01 \\
$C_{22}$  &    0.1913735962782057699247882D+01&     0.2D+01 \\
$C_{12}$  &   -0.3891044152404418071300346D+01&     0.1D+02 \\
$C_{02}$  &    0.8591820485935177732931007D+01&     0.4D+02 \\
$C_{23}$  &   -0.6438745411438708877291814D+01&     0.8D+01 \\
$C_{13}$  &    0.3104626929937093328313232D+01&     0.2D+02 \\
$C_{03}$  &   -0.1401013718110480637040870D+02&     0.4D+02 \\

\hline
\end{tabular}
\label{tab:c11}
\end{table}

\end{document}